\def\doublespaced{\baselineskip=\normalbaselineskip\multiply
    \baselineskip by 175\divide\baselineskip by 100}
\newcommand{\beq}{\begin{equation}}
\newcommand{\eeq}{\end{equation}}

\input{epsf.sty}
\documentstyle[12pt]{article}
\begin{document}
\begin{flushright}
MSUHEP-51116 \\
hep-ph/9511337
\end{flushright}
\vspace{10 mm}
\LARGE
\begin{center}
{Top-Charm Strong Flavor-Changing
Neutral Currents at the Tevatron}\\
\vspace{15 mm}
\large
 {Ehab Malkawi and Tim Tait }\\
\small
\vspace{3 mm}
Department of Physics and Astronomy,
Michigan State University \\
East Lansing, MI 48824, USA\\
\vspace{5 mm}
\large
{October 1995}
\vspace{10 mm}
\end{center}
\begin{abstract}
The possibility of an anomalous coupling
between the top and charm quarks
and the gluon field is explored in a
model-independent way using an
effective Lagrangian that is
gauge-invariant under a non-linear realization of
SU(3)$_{C} \times$ SU(2$)_{L} \times$ U(1$)_{Y}$.  Even for the
current 200 $pb^{-1}$ of
integrated luminosity at the Tevatron,
the new physics scale that strongly modifies the
coupling of t-c-g must be larger than about 2.5 TeV  if no signal
is found within a 3$\sigma$ confidence limit.  For 1 $fb^{-1}$
of data, this constraint can be pushed up to 3.8 TeV.
 \end{abstract}
\normalsize
\doublespaced
\newpage
\section{Introduction}\label{sec:intro}

\indent

With the discovery of the top quark at Fermilab by the D\O $\;$ and CDF
collaborations, it has become natural to study its interactions with
the gauge bosons.  The Standard Model (SM) completely
predicts how the top
quark should behave under these interactions, so any deviation from this
behavior would provide us with a probe of new physics beyond the SM.

The top ($t$) quark is very heavy, about 35 times that of the next
heaviest quark, the top's weak partner,
the bottom ($b$).  For this reason it is a likely place to search
for new physics.  If new physics is found in the top quark sector,
it is possible that this new physics could explain why the
top is so heavy and how its mass is generated.
This could in turn provide us
with clues as to how the other quark masses arise
(a question the SM makes
no effort to address).  
Perhaps there is new physics specific to the third family, physics which
can explain why the top, bottom, and tau lepton are so much heavier than
their first and second family counter-parts.  Or there could be new
interactions that are not really involved in producing the large masses, but
coupling more significantly to particles with large mass, and thus can be
detected by studying the top, while
only affecting the other quarks insignificantly.

The top mass is of the order of the Electro-Weak Symmetry Breaking (EWSB) 
scale $v= $ 246 GeV, and
thus provides a probe of the physics associated with the generation of the
masses for the weak gauge bosons.  The Higgs mechanism
of the SM requires a
neutral scalar particle (the Higgs boson) which
has yet to be directly detected
experimentally, but could affect (although marginally)
low energy experimental results through loop effects.
If the Higgs mechanism with Yukawa interactions
is responsible for the generation of the fermion masses,
the Higgs boson should have a coupling with the
top quark of the order of $m_{t}/v$,
and thus interactions involving the top quark may
provide a probe of the Higgs physics.

The SM does not contain tree-level flavor-changing
neutral currents (FCNC),
though they can occur at higher order through radiative corrections.
Because of the loop suppression, these
SM effects will be small, and so large
FCNC provide a window into physics beyond the SM.  In this paper, we are
specifically interested in the possibility of a top-charm-gluon ($t$-$c$-$g$)
coupling.  As we have explained above,
it is natural to look to the top quark
as a window to new physics.  If this new physics may participate in
the generation of fermion masses, it is
reasonable to assume that the coupling of the top
to the other up-type quarks should be proportional to $\sqrt{m_{t}m_{c(u)}}$
\cite{Fritzsch}.  This leads to the conclusion that the top-charm coupling is
more likely to lead to measurable effects than the top-up coupling.
A study similar to ours of
the $t$-$c$-$Z$ anomalous coupling can be found in \cite{tcz,tev2000}.

There are a number of interesting models \cite{h0tc,tc2}
in the literature that can
produce this kind of anomalous coupling, and for this reason it is
useful to study this kind of interaction experimentally in order to constrain
parameters in these theories.  In this paper we study
the $t$-$c$-$g$ vertex in an
effective Lagrangian, model-independent way.  We show how it is possible
to use single-top production data from the Tevatron to constrain the
$t$-$c$-$g$ coupling.  

\section{Theory}\label{sec:theory}

\indent

To incorporate new physics, we consider an effective
Lagrangian,
\beq
{\cal{L}}_{eff} = {\cal{L}}_{0} + {\cal{L}}_{1},
\eeq
where ${\cal{L}}_{1}$ contains operators of dimension higher than
four, multiplied by coefficients with appropriate dimensions of mass to
insure that the dimension of the Lagrangian as a whole remains four.
It is including terms of this kind that leads us to call this an "effective
Lagrangian"; since the resultant theory is not valid to an arbitrarily
high energy scale, it is not a fundamental physical theory.
Instead, it represents
the theory that is "effective" at a lower energy scale where the energy is
too low to allow us to see the full details of the underlying physics.  The
coefficients with dimensions of mass in front of the effective terms
characterize the mass scale at which new physics must enter the theory if any
non-SM effect is to be found.  In our case, since we wish to consider
the possibility of a flavor-changing gluonic current, ${\cal{L}}_{0}$
will be the QCD Lagrangian,
\beq
 {\cal{L}}_{0} = -\frac{1}{4}G^{a}_{\mu \nu}G^{a \mu \nu}
+ \overline{q}\; i\gamma^{\mu} D_{\mu}q
- m_{q}\overline{q}q.
\eeq
where $D_{\mu} = \partial_{\mu} - ig_{s}\frac{\lambda^{a}}{2}G^{a}_{\mu}$,
and $G^{a \mu \nu}$ is the usual gluon field strength tensor.

We are interested in exploring the possibility that the gluon current can
couple the top and charm quarks at tree level in an
effective theory valid up to a scale
of $A$.
${\cal{L}}_{1}$ must
be constructed in such a way as to accomplish this, respecting
the SU(3$)_{C}$ gauge
invariance of QCD.  Our effective theory must contain a cut-off
mass scale, $A$ , that is
appreciably larger than the energy scale at which we do calculations, for as
stated above, it is only in this region that the effective theory is valid.
Since any higher dimension operators we introduce will be suppressed by a
power
of the cut-off mass that fixes the over-all dimension of the term at four,
we expect that the effective operators of lower dimension should be more
important than the higher dimension ones.
The lowest dimension effective operator we can add to
produce a $t$-$c$-$g$  coupling is dimension five.
It is given by:
\beq
{\cal{L}}_{1} = \frac{g_{s}\kappa_{R}}{A}
[ \;\overline{t_{R}}\frac{\lambda^{a}}{2}G^{a}_{\mu \nu}
\sigma^{\mu \nu}c_{L} + H. c. ] +\frac{g_{s}\kappa_{L}}{A}
[ \;\overline{t_{L}}\frac{\lambda^{a}}{2}G^{a}_{\mu \nu}
\sigma^{\mu \nu}c_{R} + H. c. ],
\eeq
where $A$ represents a mass cut-off scale above which our effective
theory breaks down, and $\kappa_{L(R)}$ are dimensionless parameters that
relate the
strength of the "new coupling" to the strong coupling constant $g_{s}$
for left (right)-handed top quarks.
If the imaginary part of $\kappa_{L} \kappa_{R}^{*}$ does not equal
zero,
this interaction will
violate CP conservation.  For simplicity, we will restrict our
study to the case where $\kappa_{L(R)}$ is real, so CP is conserved.

 ${\cal{L}}_{1}$ is invariant under
local SU(3$)_{C}$ gauge transformations.  It is also invariant under a
nonlinear realization of the broken symmetry SU(2$)_{L} \times$ U(1$)_{Y}
\rightarrow$ U(1$)_{EM}$, as in the
Chiral Lagrangian \cite{chiralL0,chiralL,chiralL2}.  An
example of this is provided in \cite{chiralL}
where the SU(2$)_{L} \times$ U(1$)_{Y}$
symmetry is realized in such a way that the fermion fields transform as their 
U(1$)_{EM}$ charges.  It is clear that ${\cal{L}}_{1} $
is invariant under this realization
of SU(2$)_{L} \times$ U(1$)_{Y}$ because the
top and charm have the same electric
charge and the gluon field is an Electro-Weak singlet.

Once the terms in the Lagrangian are specified, one can
derive the Feynman rules corresponding to the vertices in the theory.
The form of the gluon field strength tensor will produce two different types
of vertices (for each $\kappa_{L(R)}$) -- a four point coupling of two
gluons and the top and charm quarks, and a
three point vertex coupling of one
gluon line and the top and charm
quarks\footnote{In Appendix A we tabulate the
Feynman diagrams resulting from ${\cal{L}}_{1}$ and QCD,
including the relevant
vertex factors.}

In order to study the $t$-$c$-$g$ couplings
introduced by ${\cal{L}}_{1}$ and to determine
the minimum energy scale $A$ that would contribute to these couplings,
we consider the
production of top-charm at a hadron collider.
There are seven tree level
diagrams (for each top quark helicity) that contribute to $t\overline{c}$,
production, as shown in Figure 1.  As we will argue below in Sec. 3, it is useful
to require that the
invariant mass $M_{t\overline{c}} \geq$ 300 GeV to separate the signal from
the background.
By energy conservation, this is the same
as requiring the invariant mass of the incoming
partons $\sqrt{\hat{s}} \geq$ 300 GeV.
As a result, the
$q\overline{q}$ annihilation diagram dominates the diagrams that have gluons
as incoming partons.  This is because the parton distribution
functions\footnote{The CTEQ2 leading order fit parton distribution functions
are used in all calculations.} are
such that for large  $\sqrt{\hat{s}} $, most of the momentum is carried by the
quarks; the probability of observing a gluon carrying a momentum fraction 
in the range we are interested in ($x = \sqrt{\hat{s}}/\sqrt{s}  \geq$ 0.15) is
much smaller than that of a quark
carrying the same momentum.  The $gg$ luminosity
is about 0.20 times the $q\overline{q}$ luminosity.  Therefore,
for $\sqrt{\hat{s}} \geq$ 300 GeV,
we can approximate the whole process by the single $q\overline{q}$
annihilation diagram.

Considering for the moment just the $t\overline{c}$ production
without any decays, we find that the
constituent cross section after
integrating over the final state phase space is given by
\beq
 \hat{\sigma}(q\overline{q} \rightarrow t\overline{c}) =
\frac{g_{s}^{4}(\kappa_{L}^{2} + \kappa_{R}^{2})}{27\pi A^{2}}\frac{(\hat{s}-m_{t}^{2})^{2}}{\hat{s}^{3}}(\hat{s}
 + 2m_{t}^{2}).
\eeq
Note that in the high energy limit (i.e. for
large $\sqrt{\hat{s}} $), $\hat{\sigma}$ is a constant
and does not have the usual functional form of  $1/\hat{s}$.
For simplicity, we do not study  $\kappa_{L(R)}$ separately in this work; here
we are interested in the energy scale at which new physics largely
modifies the $t$-$c$-$g$ coupling.  In the rest of this paper,
we restrict ourselves to
$\kappa_{L}$ = $\kappa_{R}$ (a vector-like coupling)\footnote{In principle, one 
could look at the angular distribution of the top quark's decay products to
determine the coefficients $\kappa_{L}$  and $\kappa_{R}$ for a given A.}.
This leads us to define,
\beq
\kappa^{2} = \kappa_{L}^{2} + \kappa_{R}^{2}.
\eeq
The constituent
cross section contains $\kappa$ and $A$ only in the combination
$\kappa / A$, thus it is convenient to define,
\beq
\Lambda = \frac{A}{\kappa},
\eeq
so that $\kappa$
has been absorbed.  This leaves us with one parameter to compare
with experimental data.  For $|\kappa| \sim {\cal{O}}(1)$,
$\Lambda$ is about the same as
the cut-off scale
described above, and characterizes the effective strength of the new
couplings.

One could also try to learn the theory's parameters
by considering the decay $t \rightarrow cg$, but this is not very well suited
to a hadron collider because of potentially large backgrounds from which the
signal events cannot easily be distinguished.  However, for completeness,
we present the decay width $t \rightarrow c g$, in the rest
frame of the top quark,
\beq
\Gamma( t \rightarrow c g) = \frac{8 \alpha_{s} m_{t}^{3}}{3
\Lambda^{2}}.
\eeq
The dependence on $m_{t}$ of this result may easily be understood from
dimensional arguments.
Because the $t$-$c$-$g$ coupling contains the factor $\Lambda^{-1}$,
the width must be proportional to $\Lambda^{-2}$.  If the charm
quark mass is taken to be zero, the only energy scale is $m_{t}$ and
since the width must have the dimension of energy, it must be
proportional to $m_{t}^{3}$.
In order to use this information to determine $\Lambda$, we find
the  ratio defined as,
\beq
R_{tcg} = \frac{\Gamma( t \rightarrow c g)}{\Gamma (t \rightarrow W^{+} b)},
\eeq
where,
\beq
\Gamma (t \rightarrow W^{+} b) = \frac{G_{F} \;m_{t}^{3}}{\sqrt{2} \;16\pi}
\left[1 - \frac{m_{W}^{2}}{m_{t}^{2}} \right] ^{2}
\left[1 + 2 \frac{m_{W}^{2}}{m_{t}^{2}}\right].
\eeq
is the decay width $\Gamma(t \rightarrow W^{+} b)$ \cite{Yuan}.  Combining
these two equations to find $R_{tcg}$, we find,
\beq
R_{tcg} = \frac{\sqrt{2} \;64 \alpha_{s} \pi m_{t}^{2}}{3 \Lambda^{2} G_{F}
\left[1 - \frac{m_{W}^{2}}{m_{t}^{2}} \right] ^{2}
\left[1 + 2 \frac{m_{W}^{2}}{m_{t}^{2}}\right] }.
\eeq
We will discuss the implications of this result in Sec. 5.

\section{Signal and Background}\label{sec:sigbac}

\indent

In this section we discuss the detection efficiency
of the signal events at the upgraded Tevatron
(a 2 TeV $p\overline{p}$ collider).  The top quark
decay mode studied is $t \rightarrow b W^{+}
(\rightarrow \ell^{+}\nu)$ for $\ell = e$ or $\mu$.

Including the decay of the top quark, the process $q\overline{q}
\rightarrow t \overline{c}$ results in the final state
$Wb\overline{c}$.  Thus, the signature of this process is an energetic
charged lepton, missing $E_{T}$, a $b$-quark jet from the top decay, and
a light jet.
This signal has (neglecting all of the quark masses
except for that of the top) kinematics similar to the Standard Model process
$q\overline{q} \rightarrow W^{*} \rightarrow t \overline{b}$,
which has been studied in
Ref.\ \cite{stelzer,carlson}, including the relevant backgrounds.  In this
study we consider the following intrinsic background processes:
\begin{itemize}
    \item $t\overline{q}$ produced by $W$-gluon fusion
    \item $Wb\overline{b}$
    \item $t\overline{b} \rightarrow Wb\overline{b}$
    \item $t\overline{t} \rightarrow W^{-}W^{+}\overline{b}b$
    \item $Wjj$
\end{itemize}
The potentially large background from $Wjj$ can be reduced by
requiring a $b$-quark to be present in the final state (i.e. $b$-tagging).
The CDF collaboration has effectively
implemented this procedure by using a silicon vertex detector (SVX).  For
Run~II, the efficiency is estimated to be 60\% per $b$-jet ($p_{T}^{b}
 > 20$ GeV and $\eta_{b}$ within the SVX coverage), with a probability of
less than 1\% for a light quark or gluon jet to be mis-identified as a
$b$-jet \cite{tev2000}.  We require that only a single $b$-quark
be identified in the final state in order to accept the event, and
we have ignored the possibility of a $c$-quark being mis-tagged as
a $b$-quark.

The number of events for the signal and backgrounds mentioned
above were calculated for the
Tevatron\footnote{$\sqrt{s}$ = 2 TeV p$\overline{p}$ collider} (assuming an
integrated luminosity L = 1 $fb^{-1}$)
using the Monte Carlo program ONETOP\cite{carlson,onetop}, except for
$Wjj$ which was calculated using PAPAGENO\cite{papageno}.
The $W$-gluon fusion rate is calculated from the two-body process, and
normalized by the total rate as explained in \cite{carlson}.
The top mass is taken to be
$ m_{t} =$ 175 GeV and the masses of the lighter quarks are neglected.
The signal and background cross sections
include the decay $W \rightarrow \ell \nu $ (where
$\ell =  e, \mu$) and both the $W^{+}$
and $W^{-}$ production modes are included.
(Since the dominant $W$ decay modes
are hadronic, ignoring the possibility of
$W \rightarrow$ hadrons reduces the signal, but the high $p_{T}$
charged lepton resulting
from the leptonic decay modes provides
an excellent trigger at a hadron collider.)
For now, we choose $\Lambda$ = 2 TeV.  We will discuss how to
determine this parameter from experimental data below.

Table 1 lists the primary cuts imposed to
simulate the detector acceptance.  $p_{T}$ denotes
transverse momentum, $\eta$ denotes the pseudo-rapidity, and
$\Delta R_{j\ell} = \sqrt{\Delta \phi_{j\ell}^{2} +
\Delta \eta_{j\ell}^{2}}$, where $\Delta \phi_{j\ell}$
is the difference in the azimuthal angle $\phi$ between $j$ and $\ell$.
The  $t\overline{c}$ production
rate is not very sensitive to the pseudo-rapidity
cuts
because the decay of the heavy top quark
will generally produce $b$-jets and
leptons in the central region.  Thus, narrowing
the pseudo-rapidity range should not significantly affect the signal
rates, but will reduce the $Wb\overline{b}$ background.  Only minimal
$p_{T}$ cuts for the jets are imposed here.
In our final analysis these cuts have been chosen
specifically to enhance the signal versus the background, and will be
explained below.  The events remaining after these cuts
(not including the 60\% $b$-tagging efficiency) are shown in the
second column of Table 2.

{}From the form of the constituent cross
section given in equation (4), we see that for large
$\hat{s}$, the cross section for $t\overline{c}$
production does not fall off as $1/\hat{s}$ as it will
for most of the background processes.  This suggests
that one way to improve the signal to
background ratio is to impose a cut on $\hat{s}$,
which by energy conservation is equivalent to
imposing a cut on $M_{t\overline{c}}$.  We find
that the best result is obtained by
requiring $M_{t\overline{c}} \geq$ 300 GeV.  This presents
something of a problem because
there is no way to directly measure
$M_{t\overline{c}}$.  In order to reconstruct $M_{t\overline{c}}$ we can use
the lepton momentum and the missing
transverse momentum in the equation,
\beq
M_{W}^{2} = (p_{\ell}+p_{\nu})^{2},
\eeq
to find two solutions for the missing component (i.e. the component
along the z-axis) of $p_{\nu}$, where
$p_{\nu}$ is the neutrino 4-momentum.  The solution
which better fits the decay of an on-shell top quark,
\beq
M_{t}^{2} = (p_{\ell}+p_{\nu}+p_{b})^{2},
\eeq
is chosen\footnote{In our calculation,
the width of the top quark is included via the
Breit-Wigner prescription.  For a 175 GeV top
quark, the SM width is about 1.5 GeV.}.
The other solution is discarded.  For the signal, there is only
one $b$-jet, so we can do this without any problem.  However, for most
of the background
processes, there is more than one $b$-jet, and it is impossible to
know {\em a priori} which one should be associated with the top decay.
To handle this, we randomly decide which $b$-jet to use for those processes
which contain more than one $b$-jet in the final state.
In order to allow for an off-shell $W$, we iterate this
procedure three times, generating a different Breit-Wigner mass for the $W$
each time, and keep the first result in which a solution exists.
Once we have determined the four-momenta $p_{\nu}$, $p_{\ell}$
and $p_{c}$ it is simple to use energy conservation to reconstruct
the invariant mass $M_{t\overline{c}}$.  We impose the cut
$M_{t\overline{c}} \geq$ 300 GeV on the
signal and backgrounds, and the results
are displayed in the third column of Table 2.

The $t\overline{t}$ background is
reduced by rejecting events that contain
evidence of the decay of an
additional $W$ boson.  Specifically, events
containing an additional lepton with $p_{T}^{\ell} \geq$ 20 GeV and
$|\eta_{\ell}| \leq$ 2.5, or additional distinguishable jets with 
$p_{T}^{j} > $ 20 GeV and $|\eta_{j}| \leq $ 3.5
are rejected (two parton jets are
considered distinguishable if $\Delta R_{jj}
\geq $ 0.4).
This results in a drastic reduction
of the $t\overline{t}$ background, leaving
a very small number of events from the
dileptonic decay mode of $t\overline{t}$.
The effects of these cuts have
been taken into account in the second column
of Table 2\footnote{To this order, they do not have any
effect on the other processes listed in Table 2.}.

In order to suppress the background
from $Wb\overline{b}$, we can make use
of the fact that a $b$-quark produced
from a top decay can be expected to
have a large $p_{T}$ due to the
heavy top mass (a typical value is about one third
of the top quark mass).  The $b\overline{b}$ jets from
the $Wb\overline{b}$ background
are produced by the decay of a virtual gluon,
and should have a momentum distribution that is more evenly
distributed.  We found that the cut
$p_{T}^{b} \geq$ 30 GeV reduces
the background and signal in such a way as to
provide the best enhancement of the significance of the signal.
The effects of this cut can be
seen in the fourth column of Table 2.

It is also desirable to impose a
cut on $p_{T}^{c}$.  Because the signal process
occurs in the s-channel, and the $W$-gluon fusion background
is t-channel, the distribution of the $p_{T}$ of the signal
$c$-jet should peak at a higher value than the light jet from $W$-gluon
fusion.  Adjusting the cut to find the value that provides the best enhancement
of the significance of the signal leads
us to $p_{T}^{c} \geq$ 30 GeV.  As can be seen in the
fifth column of Table 2, this cut has virtually no effect on the signal,
while slightly reducing the $W$-gluon
fusion and $Wb\overline{b}$ backgrounds.

\section{Determination of $\Lambda$}\label{sec:detlam}

\indent

In order to set bounds on the parameter $\Lambda$,
it is necessary to look for
evidence of a signal distinguishable from
background fluctuations.  We require
a 3$\sigma$ effect as our criterion
for judging the signal to
be distinguishable from a background fluctuation,
that is, we require the probability
for the background to fluctuate up to the
observed level to be less than 0.27\%.
The number of signal events, $N_{S}$,
has a simple dependence on the integrated luminosity and $\Lambda$,
\beq
N_{S} = \frac{{\rm L}\alpha}{\Lambda^{2}} .
\eeq
The coefficient $\alpha$, which characterizes the
acceptance of signal events multiplied by the
signal cross section after extracting the factor
$1/\Lambda^{2}$, is determined by our Monte Carlo study at
$\Lambda =$ 2 TeV and L = 1 $fb^{-1}$ to be
$\alpha$ = 616 ${\rm TeV^{2}}fb^{-1}$.
The number of background events, $N_{B}$, is just the integrated
luminosity multiplied by the total background cross sections.  Once $N_{B}$
has been determined for a given L, we can determine the $N_{S}$ that will
provide a 3$\sigma$ effect using Gaussian statistics (since $N_{B}$ and
$N_{S} \geq$  10).  From
$N_{S}$ one can then determine the minimum $\Lambda$ such that a 3$\sigma$
effect is not observed at the integrated luminosity of interest.  In Table 3
we list the constraints on $\Lambda_{min}$, the
minimum value of $\Lambda$ for which new
physics will show up in $t$-$c$-$g$ couplings,
for several different integrated
luminosities provided a 3$\sigma$ effect is not observed at the Tevatron.

We expect that because we are examining the high invariant mass
$M_{t\overline{c}}$ region, the theoretical uncertainties due to the choice
of the scale, $Q^2$, in $\alpha_{S}$, and in the parton distribution functions 
used for calculations, should be small
(Note that in the relatively large $x$ region, the parton distribution functions
are well determined by deep inelastic scattering data.)
In order to quantify
this effect we have examined the signal cross section using $Q^2 = \sqrt{\hat{s}}$
and $Q^2 = \sqrt{m_{t}^2+{p_{T}^{t}}^2}$.  We find that the cross sections in these
two cases differ by about 17\%, which can affect our determined
$\Lambda$ by about 10\%.

\section{Discussion and Conclusions}\label{sec:concl}

\indent

We have shown that it is possible to introduce an operator of
dimension five into the QCD Lagrangian that couples the
top and charm quarks to the gluon field.  This
operator can be constructed to
respect the SU(3$)_{C}$ local gauge
symmetry of QCD, and the broken symmetry
SU(2$)_{L} \times$U(1$)_{Y} \rightarrow$ U(1$)_{EM}$
realized non-linearly, as in
the Chiral Lagrangian.  Because the operator is dimension five, it is
divided by a parameter with the dimension of mass,
$\Lambda$, that characterizes the
energy scale at which new physics will modify
the $t$-$c$-$g$ interactions.

We have completed a Monte Carlo study of the process
$q\overline{q} \rightarrow t\overline{c}$
and the SM backgrounds present at the Tevatron to show how to
enhance the ratio of the signal to the
background to improve the limits one may set on the
minimum value of $\Lambda$, $\Lambda_{min}$.
As shown in Table 3, our
results indicate that  even with an integrated
luminosity of 200 $pb^{-1}$, the lack
of a 3$\sigma$ signal at the Tevatron requires $\Lambda \geq$ 2.5 TeV
(assuming a $b$-tagging efficiency of 60\%\footnote{ A 30\%
$b$-tagging efficiency will reduce the constraint
on $\Lambda_{min}$ to about $\Lambda_{min} \geq$ 1.7 TeV.} ).
Assuming $\Lambda =$ 2.5 TeV, there would
be 19 signal events and 41 background
events produced at the Tevatron with
an integrated luminosity of 200 $pb^{-1}$.
A higher integrated luminosity allows
one to set even stronger limits on $\Lambda$.
For instance, for 1 $fb^{-1}$ of data, $\Lambda \geq$ 3.8 TeV
if no signal is found at a 3$\sigma$
level.  If a signal is found, we expect at least
43 signal events with 207 background events.
From equation (10) we see that if  $\Lambda =$ 2.7 TeV,
the branching ratio $t \rightarrow c g$ should be about 13\%.  This
suggests that a detailed study of the decay of the top quark can
probe $\Lambda$, but because of large backgrounds from which
this signal cannot easily be distinguished, study of $t \overline{c}$
production should provide a better way to set experimental bounds
on $\Lambda$.
Thus, a detailed study of single-top production at the Tevatron is useful
in that it will allow one to set bounds
on the possibility of  this type of anomalous
top quark coupling.

In this work, we did not consider the other possible
single-top production processes
$q c \rightarrow q t$ , $\overline{q} c \rightarrow
t \overline{q}$ , and $g c \rightarrow g t$ as shown
in Figure 2.  We speculate that due
to the large backgrounds discussed in Sec. 3,
it is necessary to impose a large $p_{T}$
cut on the final state parton jet $q$, $\overline{q}$, or $g$.
This cut should significantly reduce these contributions to
the signal, making it difficult to observe.
Further study is needed to better understand this.

\newpage
\section{Acknowledgements}\label{sec:ackn}

\indent

The authors would like to thank C.--P. Yuan and X. Zhang
for helpful discussion.
This work was supported in part by the NSF
under grant no. PHY-9309902.

\newpage
\begin{center}
{\bf Table 1: Fundamental Kinematic Cuts} \\
\vspace{5 mm}
\begin{tabular}{|c|} \hline
$p_{T}^{j} > $ 10 GeV \\

$p_{T}^{\ell} > $ 20 GeV \\

$|\eta_{\ell}| \leq$ 2.5 \\

$|\eta_{j}| \leq $ 3.5 \\

$|\eta_{b}| \leq $ 2.0 \\

$\Delta R_{j\ell} \geq $ 0.4 \\

$\Delta R_{jj} \geq $ 0.4 \\ \hline
\end{tabular}
\end{center}
\newpage
\begin{center}
{\bf Table 2: Total number of events at the Tevatron,
including kinematic cuts,
for L = 1 $fb^{-1}$ and $\Lambda =$ 2 TeV.  The
"Fundamental Cuts" denote those in Table 1.}
\vspace{10 mm}
\\
\begin{tabular}{|c|c|c|c|c|c|} \hline
Process & Fundamental
& $M_{t\overline{c}} \geq$ 300 GeV
& $p_{T}^{b} \geq$ 30 GeV
& $p_{T}^{c} \geq$ 30 GeV & $b$-tagging \\ \hline

$t\overline{c}$ & 308 & 276 & 256 & 256 & 154 \\

$W$-g fusion & 188 & 168 & 152 & 132 & 79 \\

$Wb\overline{b}$ & 168 & 124 & 88 & 72 & 35 \\

$t\overline{b}$ & 40 & 36 & 28 & 28 & 13 \\

$t\overline{t}$ & 4 & 4 & 0 & 0 & 0 \\

$Wjj$                &     &    &    &  4000 & 80 \\

Tot. Background &  &  &  & 4232 & 207 \\ \hline
\end{tabular}
\end{center}
\newpage
\begin{center}
{\bf Table 3 : $\Lambda_{min}$ for several Integrated Luminosities}
\vspace{10 mm}
\\
\begin{tabular}{|c|c|c|c|} \hline
L & $\Lambda_{min}$ & $N_{S}$ & $N_{B}$ \\ \hline
200 $pb^{-1}$& 2.5 TeV & 19 & 41 \\

1 $fb^{-1}$ & 3.8 TeV & 43 & 207 \\

2 $fb^{-1}$ & 4.5 TeV & 61 & 414 \\

10 $fb^{-1}$ & 6.7  TeV & 136 & 2070 \\

20 $fb^{-1}$ & 8.0 TeV & 193 & 4140 \\ \hline
\end{tabular}
\end{center}

\newpage
\pagestyle{empty}
\noindent
\epsfbox[60 0 0 620]{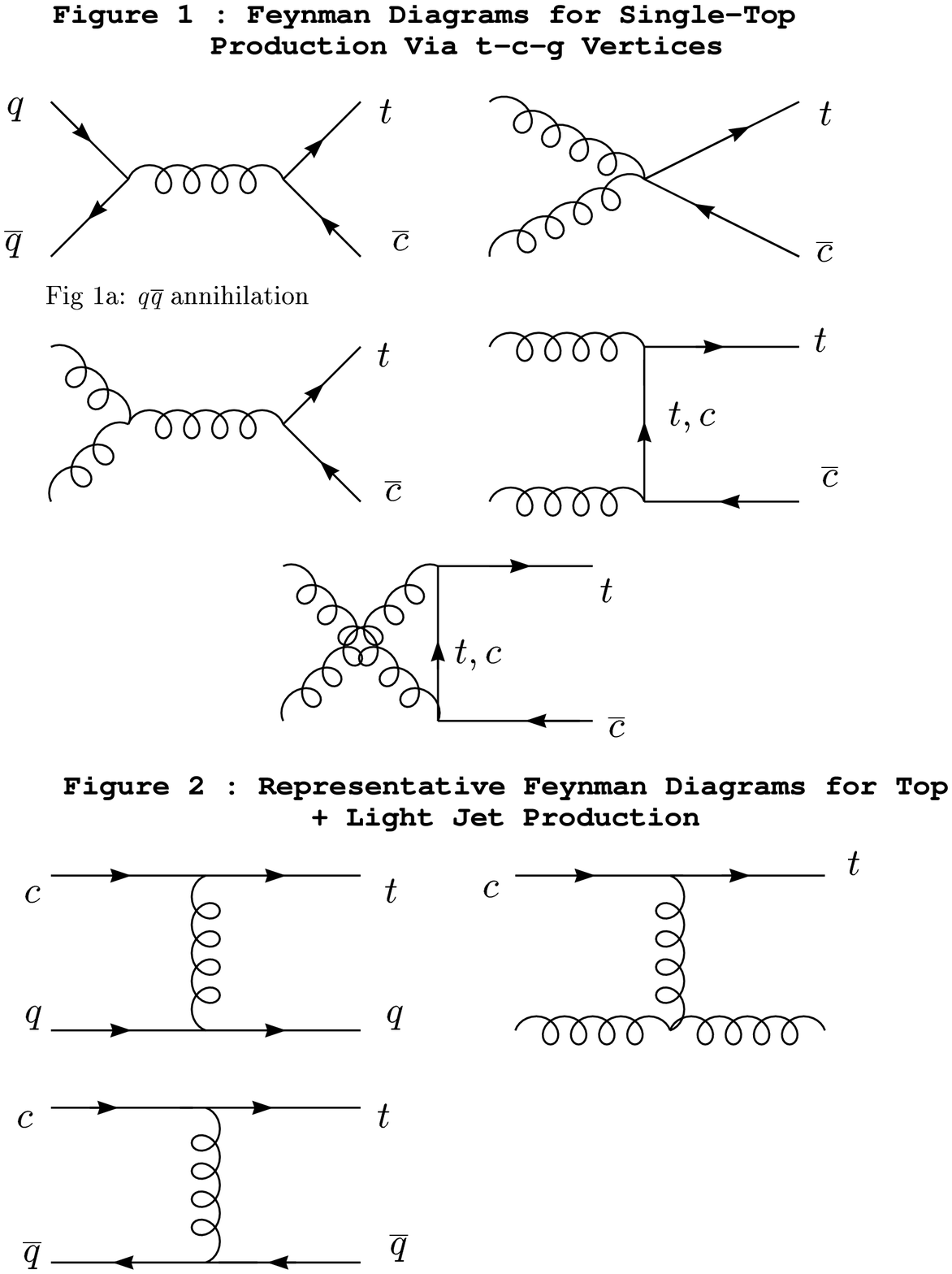}

\newpage
\pagestyle{empty}
\noindent
\epsfbox[60 0 0 650]{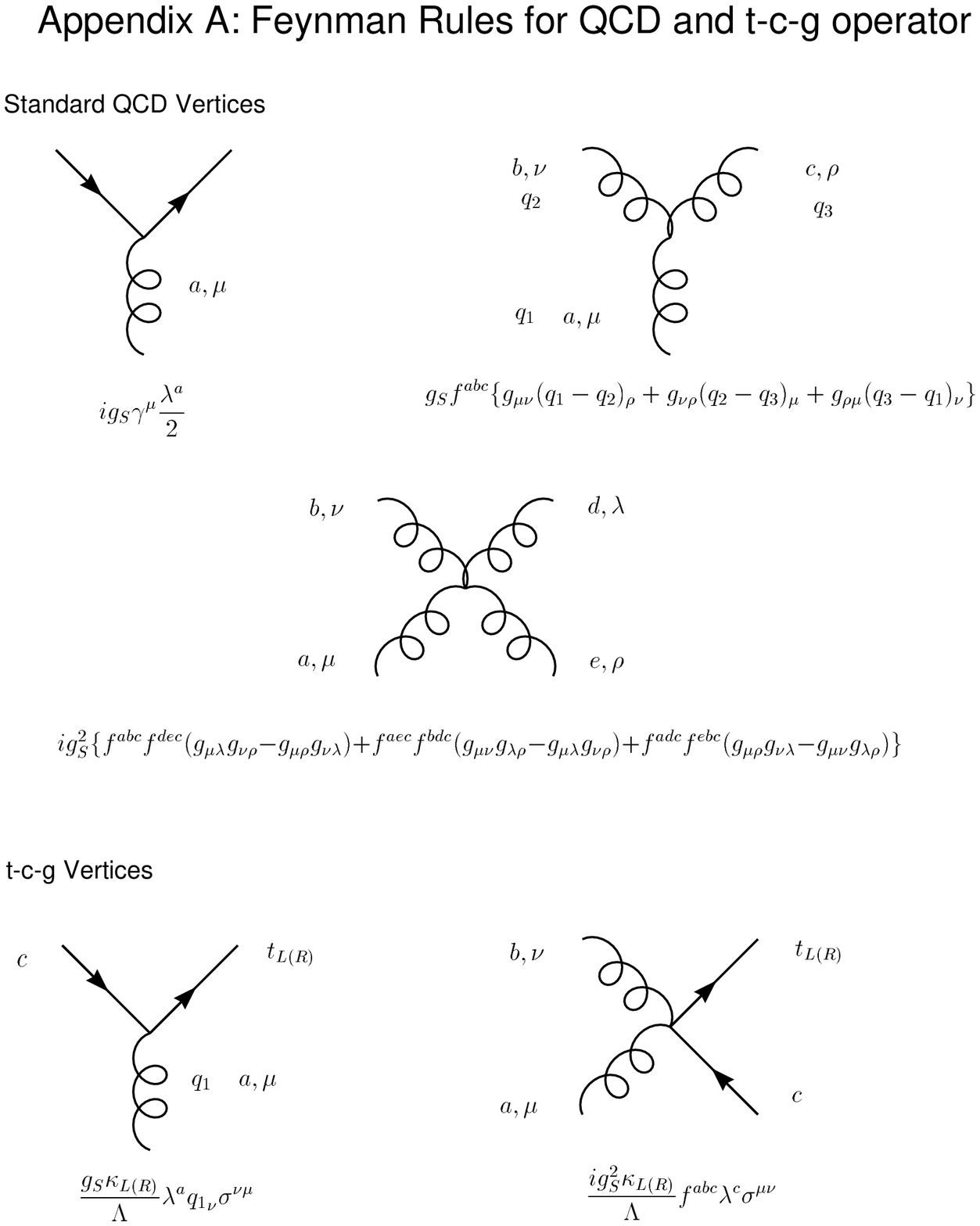}

\end{document}